\documentclass{gGAF2e}

\pdfoutput=1

\begin{document}

\jvol{00} \jnum{00} \jyear{2012} 

\markboth{\rm A. SUKHANOVSKI AND E. POPOVA}{\rm GEOPHYSICAL \&  ASTROPHYSICAL FLUID DYNAMICS}


\title{{A shallow layer laboratory model of large-scale atmospheric circulation}}

\author{Andrei Sukhanovskii${\dag}$\thanks{$^\ast$Corresponding author. Email: san@icmm.ru
\vspace{6pt} } and Elena Popova${\dag}$ \\\vspace{6pt}
Institute of Continuous Media Mechanics,  Korolyov 1, Perm, 614013, Russia\\ 
\vspace{6pt}\received{October 2022} }

\maketitle

\begin{abstract}

A new shallow layer laboratory model of global atmosphere circulation is realized.  The shallow rotating cylindrical layer of fluid with the localized heater at the bottom periphery and localized cooler in the central part of the upper boundary is considered. The rim heater imitates the equator heating and disc cooler -- the north pole cooling. The rim heater is intentionally shifted from the sidewall to decrease the influence of the no-slip vertical boundaries and provide anticyclonic-cyclonic motion in the upper layer, imitating the formation of large-scale zonal flows (easterlies and westerlies) in the low latitudes. The low-viscous silicon oil is used instead of water to avoid complex effects provided by the formation of a thin film of a surface-active substance on the open surface. The flow transforms from the Hadley-like regime to the baroclinic wave regime through transitional states. The decrease in the thermal Rossby number for the fixed value of Taylor number results in the regularization of the baroclinic waves. The main difference between presented and classical annulus configurations is the absence of the steady waves. All wave regimes, even with regular wave structures, are characterized by strong non-periodic fluctuations. The observed baroclinic wave structures are a combination of temporarily evolving different baroclinic modes. The presented model provides the formation of atmospheric-like flows, characterized by complex temporal behaviour and can serve as an efficient tool for studying different aspects of global atmospheric circulation.  


\end{abstract}

\section{Introduction}

The global atmospheric circulation plays a crucial role in weather and climate processes providing the transfer of heat and angular momentum. The structure and dynamics of the global atmospheric circulation are very complex and depend on multiple factors such as rotation, solar heating, surface topography e.t.c. Remarkable progress in the numerical modeling of global circulation of last decades is made. However, this should not be misleading, because the atmosphere is a highly non-linear system and we are far from full understanding of complex interactions and links between its different elements. The global atmospheric circulation has convective nature and its driver is a meridional temperature difference, which can variate due to different processes. For example, substantial warming amplification over the Arctic Pole results in decreasing of temperature contrast between the pole and the equator  \citep{you2021warming}. 

Decades of studies have proven that relatively simple laboratory models can help to understand the nature of complex atmospheric processes \citep{von2014modeling}. Laboratory modeling makes it possible to isolate some key processes and study them with a reproducible and controlled series of experiments.

The main laboratory model for mid-latitude circulation was developed by \cite{hide1953some} and consists of a rotating annular vessel with three concentric cylinders, where the narrow gap of outer annulus is filled with warm fluid, the inner cylinder is filled with cold fluid, and the central part is the experimental chamber filled with water. The well-controlled boundary conditions and similarity with Eady model \citep{eady1949long} help to analyze the experimental results.  This configuration has been successfully used  for many years to study various aspects related to baroclinic wave formation in mid-latitudes \citep{read2014general}.  Recent studies are focused on the influence of the mechanical and thermal inhomogeneity on the baroclinic wave formation  \citep{marshall2018experimental,marshall2020thermal}, the generation of different waves in the barostrat instability experiment \citep{rodda2018baroclinic}. First laboratory studies of mid-latitude circulation in a frame of a polar amplification scenario \citep{vincze2017temperature,wcd-3-937-2022} showed that a decrease in the meridional temperature difference slows down the east-ward propagation of the jet stream and complexifies its structure. A good agreement between the probability distribution of extreme events in a laboratory experiment and in the  atmospheric case, proves the usefulness of laboratory modeling for studying the dependence of the probability of extreme events on climate change \citep{harlander2022probability}. 

However, the real atmosphere is rather characterized by effective heating near the ground in the tropics and cooling in the upper layer in the polar region, which limits the applicability of classical annular configuration with vertical isothermal boundaries and motivate realization of a new laboratory configuration for the modeling of atmospheric circulation \citep{scolan2017rotating}. The horizontal and vertical displacements of heat and cold sources may have a strong influence on the flow structure and dynamics. Indeed, the large-scale flows in the alternative configuration, so-called dishpan experiments, with a cylindrical vessel with the rim heating at the bottom periphery and cooling at the center \citep{fultz1959studies} were very similar to the typical atmospheric flows but less steady and regular than in the classical annular configuration. The complexity and irregularity of flows in a dishpan configuration was the main reason why a rotating annular vessel with isothermal sidewalls was chosen as a basic model of large-scale atmospheric circulation. 

Only recently, first steps toward a more complex atmospheric-like model \citep{scolan2017rotating}, were done. In a new proposed model the convection in a rotating annular vessel is driven by local heating at the bottom and cooling at the center of the upper boundary. It is shown that the flows in this configuration exhibit more spatio-temporal complexity, including  coexistence and interaction between free convection and baroclinic wave modes. These features were not observed in the classical configuration and validate the new model as a tool for studying the atmosphere-like dynamics in a more realistic framework. 

Another important parameter of global atmospheric circulation modeling is the aspect ratio. The ratio of the characteristic vertical scale (about 20 kilometers) to the horizontal one (more than one thousand kilometers) for large-scale atmospheric flows is very small, which is not the case for the experimental models. For the classical annulus configuration, the aspect ratio is usually larger than 1 \citep{read2014general} and for the model of \cite{scolan2017rotating} it is about  0.65. Even for the steady Hadley-like regime, when the one-cell meridional circulation is formed, despite qualitative similarity, there are remarkable quantitative differences in the flow characteristics for the deep layer  \citep{hignett1981rotating,read1986} and shallow layer \citep{batalov2010}. Transition from the steady Hadley-like regime to the more complex regimes with baroclinic and free convection modes may result in more serious differences between deep and shallow layers. Here, we present a first series of experiments using a shallow layer model of large-scale atmospheric circulation.

\section{Experimental set-up}

\begin{figure*}
\center{\includegraphics[width=0.6\linewidth]{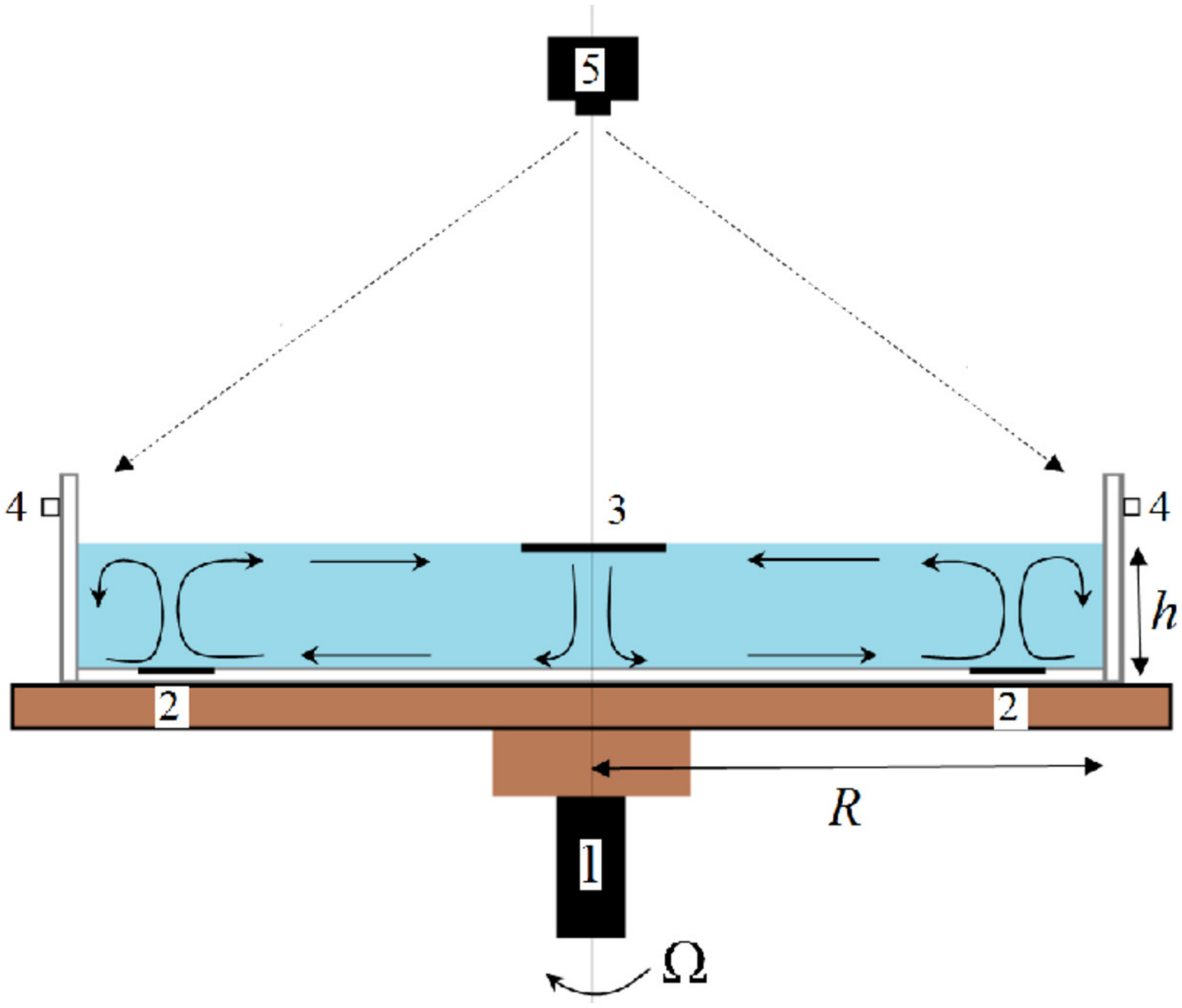}} a)\\
\center{\includegraphics[width=0.6\linewidth]{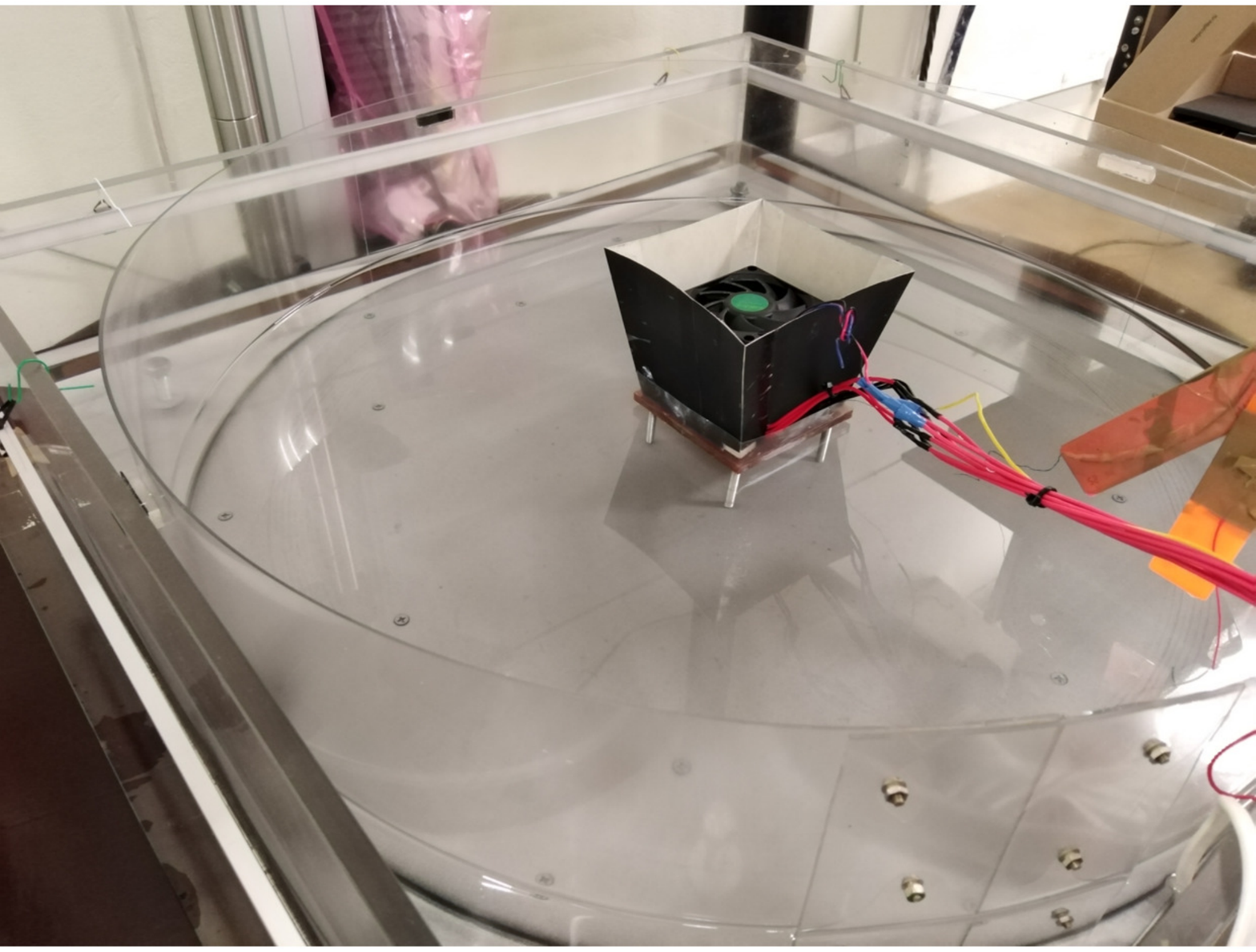}} b)\\
\caption{(a) Scheme of the laboratory model, 1 - rotating table, 2 -  rim heater, 3  - cooler, 4 - LED illumination, 5 - CCD camera; (b) Photo of the experimental set-up. (Colour online)}
\label{Fig1}
\end{figure*}

The principal scheme of the laboratory model of the general atmosphere circulation is shown in Fig.~\ref{Fig1}a. The shallow rotating cylindrical layer of fluid with the localized heater at the bottom periphery and localized cooler in the central part of the upper boundary is considered. The rim heater imitates the equator heating and disc cooler -- the north pole cooling. The rim heater is intentionally shifted from the sidewall to decrease the influence of the no-slip vertical boundaries and provide anticyclonic-cyclonic motion in the upper layer, imitating the formation of large-scale zonal flows (easterlies and westerlies) in the low latitudes.

The experimental model is a rectangular tank of a square cross-section with a side $L = 700$ mm, and height $H = 200$ mm (Fig.~\ref{Fig1}). The sidewall and bottom are made of Plexiglas with a thickness  20 mm. For realization of cylindrical layer the Plexiglas cylinder with 3 mm wall and  diameter $D = 690$ mm is inserted into the tank. There is no cover or additional heat insulation at the sidewall. The cylindrical insert is installed on a false bottom, the upper part of which is a foil-coated textolite. By cutting out the tracks in the foil, it is possible to make heaters of different configurations. At the present case the heater is a circular stripe of width $l = 25$ mm, heated by electrical current. The distance from the cylindrical sidewall to the outer border of the heater is 40 mm.  The heater is a very thin copper foil (of about 50 micrometers) so the temperature of its surface depends strongly  on the flow structure. The heating power is controlled and kept constant during the experiment. The room temperature is kept constant by an air-conditioning system, and cooling of the fluid is provided  by the heat exchange with the surrounding air on the free surface, the central cooling system and some heat losses through the sidewall. The cooling system includes a thick (10 mm) copper disc with diameter $d=54$ mm partially inserted into the upper layer of the fluid (about 2 mm).  We chose the cooler of relatively small area to minimize the impact of the solid lid, because the friction on solid boundaries plays an important role in an angular momentum balance in rotating convection \citep{evgrafova2022angular}. The upper surface of the copper disc is cooled by a thermoelectric (Peltier) cooler. To remove heat from the hot side of the thermoelectric cooler a radiator with a forced air circulation is used. For minimization of the impact of the air circulation, the cooling system is surrounded by additional open box. The temperature of the cooler was measured by a copper--constantan thermocouple installed into the copper disc, and the cooling power is estimated using the known relation between Nusselt number and Rayleigh number. 

The experimental model is placed on a rotating horizontal table. The rotating table provides a uniform rotation in the angular velocity range $0.02\leq\rm{\Omega}\leq 0.30$ rad s$^{-1}$ (with accuracy of $\pm0.001$ rad s$^{-1}$). Silicon oil PMS-5 is used as the working fluid. In all presented experiments, the depth of the fluid layer $h$ was 30 mm and the  surface of the fluid was open. The temperature inside the fluid layer was measured at mid-height ($z = 15$ mm) and $R=180$ mm  by the copper--constantan thermocouple and used for the estimation of the mean temperature of the fluid. The main fluid properties and parameters of the experimental set-up are provided in Table~\ref{tab:exp}. The direction of rotation in all presented experiments is clockwise.

\begin{table}
  \begin{center}
\def~{\hphantom{0}}
  \begin{tabular}{lcccc}
       Fluid properties & ~~Symbol   & ~Value & ~~Units \\[3pt]
            density   & ~~$\rho$~ & ~~911 & ~~kg $\rm m^{-3}$  \\
            kinematic viscosity   & ~~$\nu$~ & ~~5.2x$10^{-6}$ & ~~$\rm m^2$ $\rm s^{-1}$ \\
           thermal diffusivity   & ~~$\kappa$~ & ~~8.3x$10^{-8}$ & ~~$\rm m^2$ $\rm s^{-1}$ \\
            thermal expansion coefficient   & ~~$\beta$~ & ~~9x$10^{-4}$ & ~~$\rm K^{-1}$ \\
              Prandtl number & ~~Pr=$\nu$ $\kappa^{-1}$~ & ~~62.7 & ~~  \\
            \\
          Experimental set-up  & ~~  & ~& ~~ \\[3pt]
           layer radius& ~~$R$~ & ~~345 & ~~mm  \\
           layer depth& ~~$h$~ & ~~30 & ~~mm  \\
           heater width& ~~$l$~ & ~~25 & ~~mm  \\
           heater radius& ~~$r_h$~ & ~~293 & ~~mm  \\
           cooler radius& ~~$r_c$~ & ~~28 & ~~mm  \\
           heating power& ~~$P_h$~ & ~~123 & ~~Wt  \\
           cooling power& ~~$P_c$~ & ~~$\approx$ 3 & ~~Wt  \\
     
  \end{tabular}
  \caption{The main fluid properties and parameters of the experimental model.}
  \label{tab:exp}
  \end{center}
\end{table}

Aluminum powder is used to visualize the flow structure in the upper layer. The illumination of the tracers is provided by LED (light-emitting diode) strip placed on the perimeter of the experimental model above the fluid layer. The aluminum flakes are oriented along the flow, so they are bright when the flow is horizontal and dark when vertical motions are dominant. This dependence of the brightness of aluminium flakes on the ratio of horizontal and vertical velocity helps to detect 3D structures. The recording was provided by 4 Mpx CCD camera Bobcat 2020. Most of the images were recorded with 1 fps.

As non-dimensional governing parameters, following \citep{scolan2017rotating} we use the thermal Rossby number ${Ro_T}$, the Taylor number $Ta$, and the Ekman number $E$ : 
\begin{align}
 {{Ro_T}} =\,&\, \frac{g \beta h \Delta T }{ \Omega^2 R^2 }, \label{Rossby}\\
{Ta} =\,&\,\frac{4\Omega^2 R^5}{h\nu^2}, \label{Reynolds}\\
 {{E}} = \,&\,\frac{\nu}{ \Omega h^2 }, \label{Prandtl}
\end{align}
where $g$ is the gravitational acceleration, $\beta$ thermal expansion coefficient, $\Delta T$ temperature difference between heater and cooler, $R$ radius of the layer, $\nu$ kinematic viscosity. Other important non-dimensional parameters, which describe conbection and the heat transfer are the Rayleigh number $Ra$ and the Nusselt number $Nu$ (ratio of the full heat flux to the conductive one):

 \begin{align}
 {Ra} =\,&\,\frac{g\beta\Delta T h^3}{\nu\kappa}, \label{Rayleigh}\\
 {{Nu}} =\,&\, \frac{f h }{ \lambda \Delta T  }, \label{Nusselt}
\end{align}

 where $\kappa$ is the thermal diffusivity, $\lambda$ is the thermal conductivity, $f$ is the heat flux.  For the calculation of the thermal Rossby number, the temperature of the heater is required. The measurement of the temperature of the thin foil is a technically complex problem. The known relation between Nusselt number and Rayleigh number $Nu=C Ra^{\alpha}$, can be used for the estimation of the mean heater temperature. In case of the developed convective regime, we can take $C=0.1$ and $\alpha=1/3$ \citep{golitsyn1979simple,evgrafova2019specifics}. After simple calculations for the fixed heat flux ($f_h=$2.45x$10^3$ Wt $\rm m^{-2}$) we obtain the mean value of the temperature difference between the fluid and the heater about $28$ K. Using the same approach, we estimate the heat power of the cooler (the temperature of the cooler is measured). For the substantial temperature difference between the cooler and the fluid (about 14 K) the cooling power in the presented experiments is approximately 3 Wt and the mean heat flux $f_c\approx$1.2x$10^3$ Wt $\rm m^{-2}$. This remarkable difference in the heating and cooling power is a consequence of the large difference in the heating and cooling areas. The cooler is substantially less than the heater (almost 19 times) and so is the total cooling power. This means that most of the cooling is provided by the heat exchange between the fluid and air on the open surface. 
 
 \section{Main results}
\subsection{Flow regimes}

For the first series of experiments we fix the heating and cooling power and change only the rotation rate.  The measurements are realized in a quasi-stationary state, when the net heat flux is approximately zero. The values of the main parameters of the experiments are presented in Table~\ref{tab:kd}.

\begin{table}
  \begin{center}
\def~{\hphantom{0}}
  \begin{tabular}{lccccc}
       Exp. & ~$\Omega$, rad $s^{-1}$ & ~$\Delta T$ & ~~$Ro_T$   & ~$Ta$  & ~~$E$  \\[3pt]
          ~~1   & ~~~~0.08   & ~~41.9 ~ & ~~14 & ~~1.7x$10^8$ & ~~0.068 \\
          ~~2   & ~~~~0.09   & ~~42.9 ~ & ~~10.8 & ~~2.3x$10^8$ & ~~0.061 \\
          ~~3   &  ~~~~0.11   & ~~44.9 ~ & ~~8.2 & ~~3.2x$10^8$ & ~~0.05 \\
          ~~4   &   ~~~~0.13   & ~~40.9 ~ & ~~5.1 & ~~4.7x$10^8$ & ~~0.042 \\
          ~~5   &   ~~~~0.17   & ~~41.9 ~ & ~~3.2 & ~~7.5x$10^8$ & ~~0.033 \\
          ~~6   &   ~~~~0.23   & ~~39.9 ~ & ~~1.6 & ~~1.4x$10^9$ & ~~0.024 \\
          ~~7   &    ~~~~0.37   & ~~39.9 ~ & ~~0.7 & ~~3.6x$10^9$ & ~~0.015 \\
          ~~8a   &    ~~~~0.48   & ~~40.9 ~ & ~~0.4 & ~~6x$10^9$ & ~~0.012 \\
           ~~8b   &    ~~~~0.48   & ~~17.1 ~ & ~~0.2 & ~~6x$10^9$ & ~~0.012 \\
            ~~8c  &    ~~~~0.48   & ~~12.1 ~ & ~~0.1 & ~~6x$10^9$ & ~~0.012 \\

  \end{tabular}
  \caption{Main parameters of experiments.}
  \label{tab:kd}
  \end{center}
\end{table}

\begin{figure*}
\includegraphics[width=0.5\linewidth]{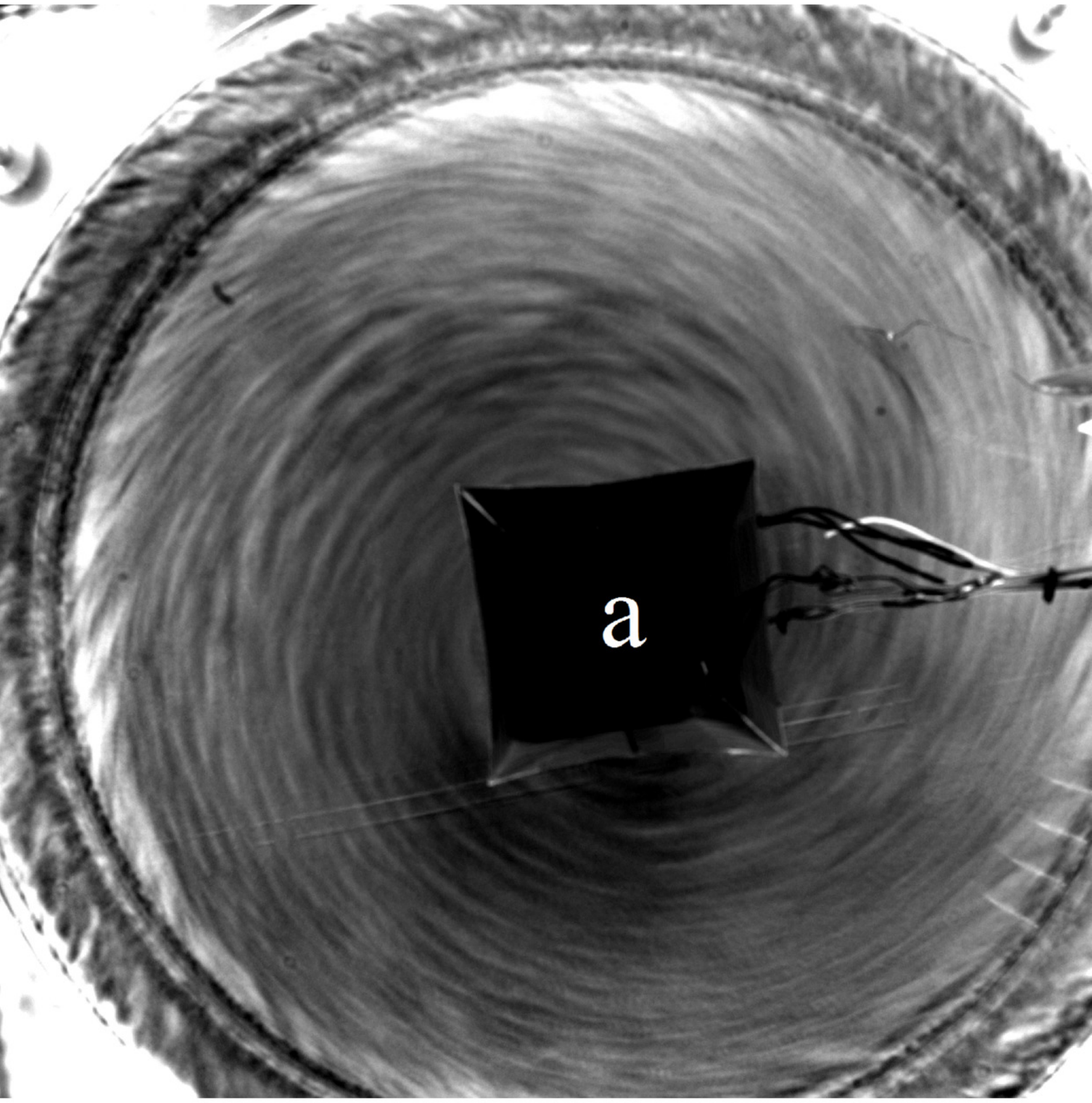} 
\includegraphics[width=0.5\linewidth]{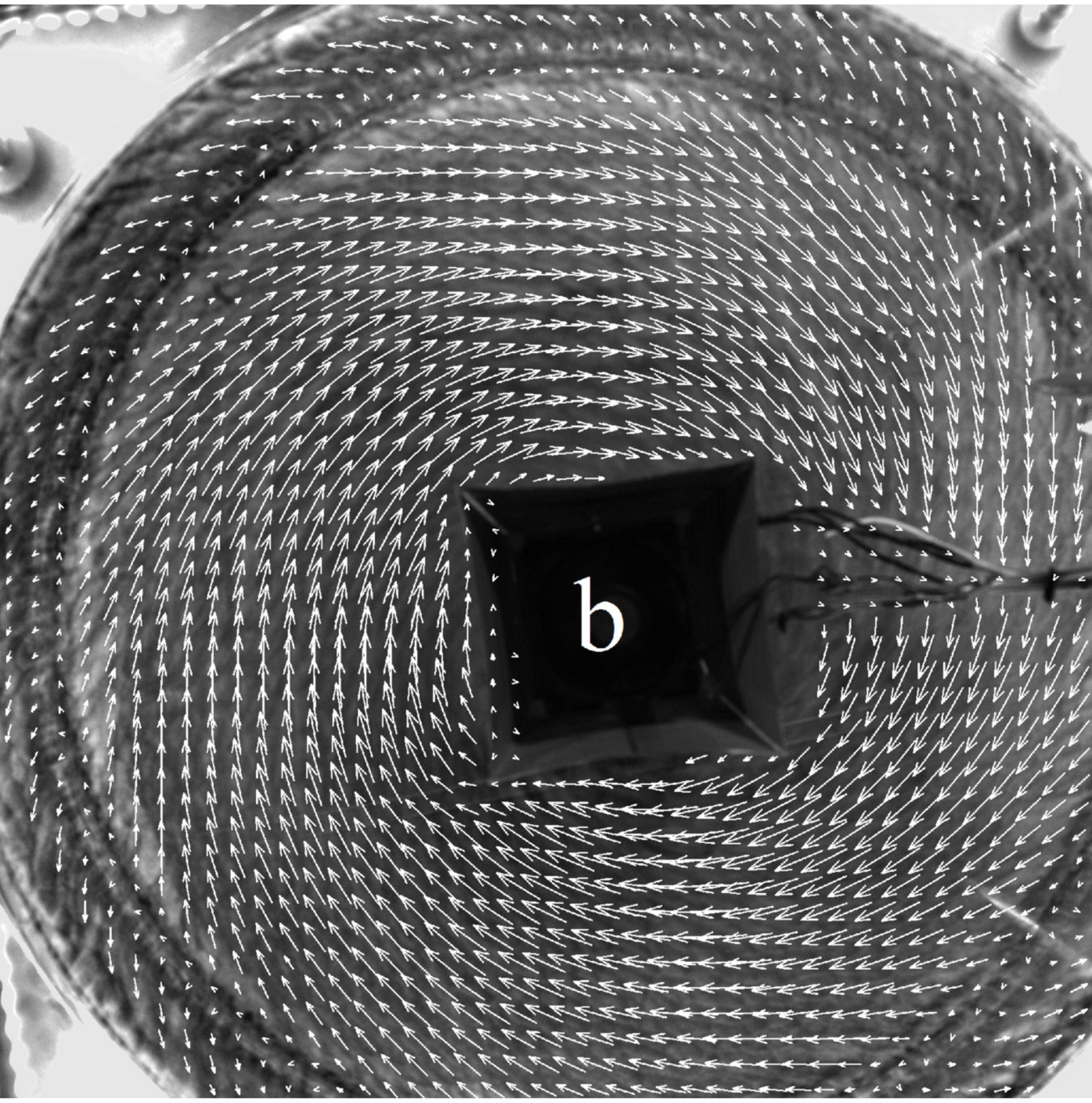}
\caption{Flow structure for the Hadley-like regime (Experiment 1a), visualization by aluminum flakes. a -- 5-frame averaging, b -- combined image with a reconstructed mean velocity field.}
\label{Fig2}
\end{figure*}

The flow structure for the relatively small angular velocity (Exp.1, Table~\ref{tab:kd}) is shown in Fig.~\ref{Fig2}. Please note, that the size of the cooler, which is in a direct contact with a fluid, is substantially less than visible part of the cooling system, including the radiator and open box. The circular stripe (Fig.~\ref{Fig2}a) visible in the periphery is a heater, which provides intensive convective flow, consisting of multiple plumes. In agreement with a scheme presented in Fig.~\ref{Fig1}a, an ascending convective flow over the heater provides formation of the convergent and divergent radial flows in the upper layer. Radial transport of angular momentum provides an anticyclonic circulation (easterlies) near the sidewall and cyclonic circulation (westerlies) at radii smaller than the radius of the heater. The observed cyclonic part of the flow is nearly axisymmetric and corresponds to the Hadley-like regime\citep{fultz1959studies,batalov2010,scolan2017rotating}.  The complex pattern formed by aluminum flakes can be used for reconstruction of velocity field by PIV (Particle Image Velocimetry) technique \citep{raffel1998particle}. The mean velocity field is presented in Fig.~\ref{Fig2}b. There is an anticyclonic belt near the sidewall and substantially more intensive cyclonic circulation in the central part.

\begin{figure*}
\includegraphics[width=0.33\linewidth]{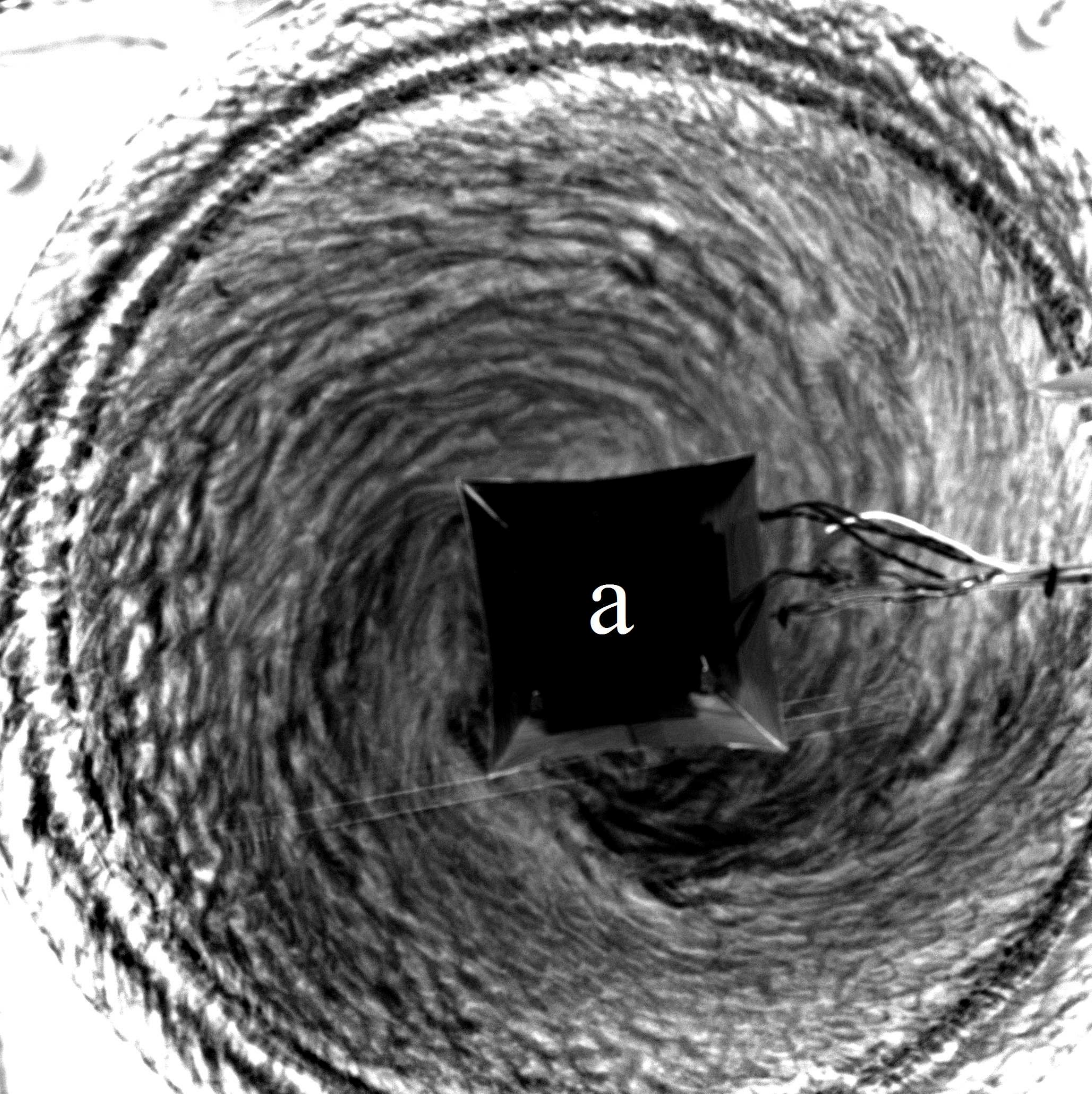} 
\includegraphics[width=0.33\linewidth]{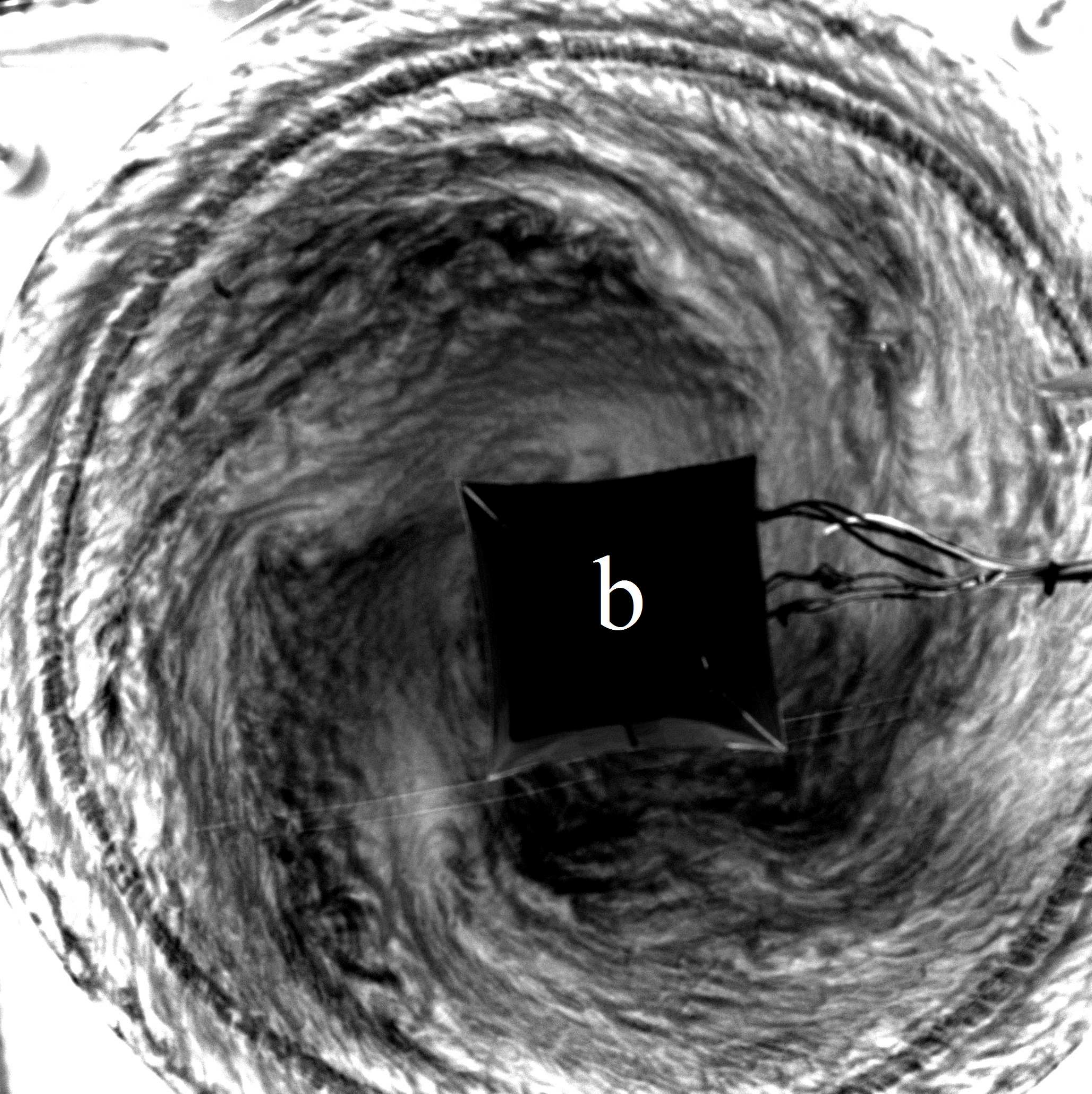}
\includegraphics[width=0.33\linewidth]{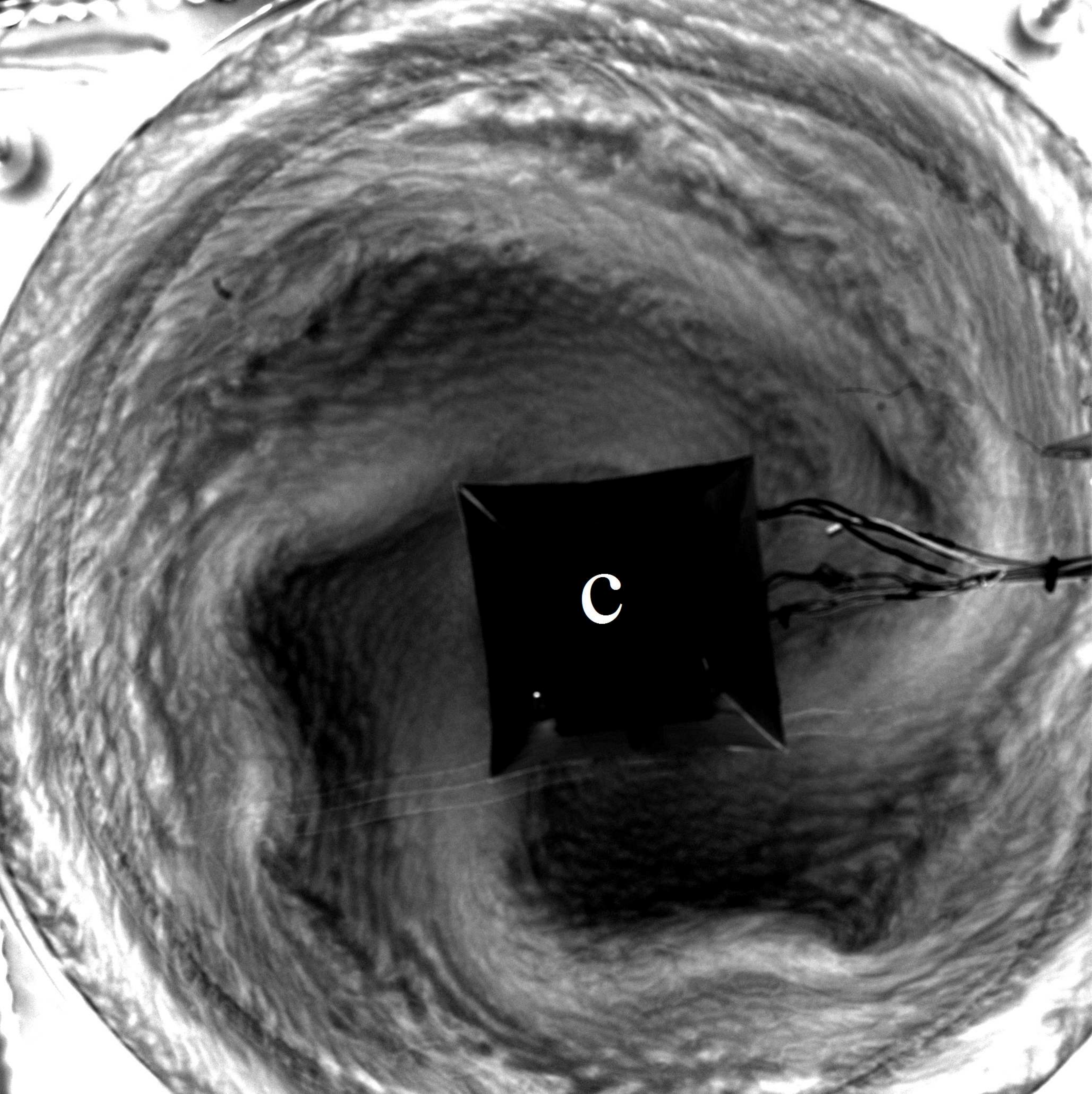} 

\caption{Instanteneous images of the flow structure for different rotation rate. a -- Exp.3, b -- Exp.4, c -- Exp.6 (Table~\ref{tab:kd}).}
\label{Fig3}
\end{figure*}

\begin{figure*}
\includegraphics[width=0.33\linewidth]{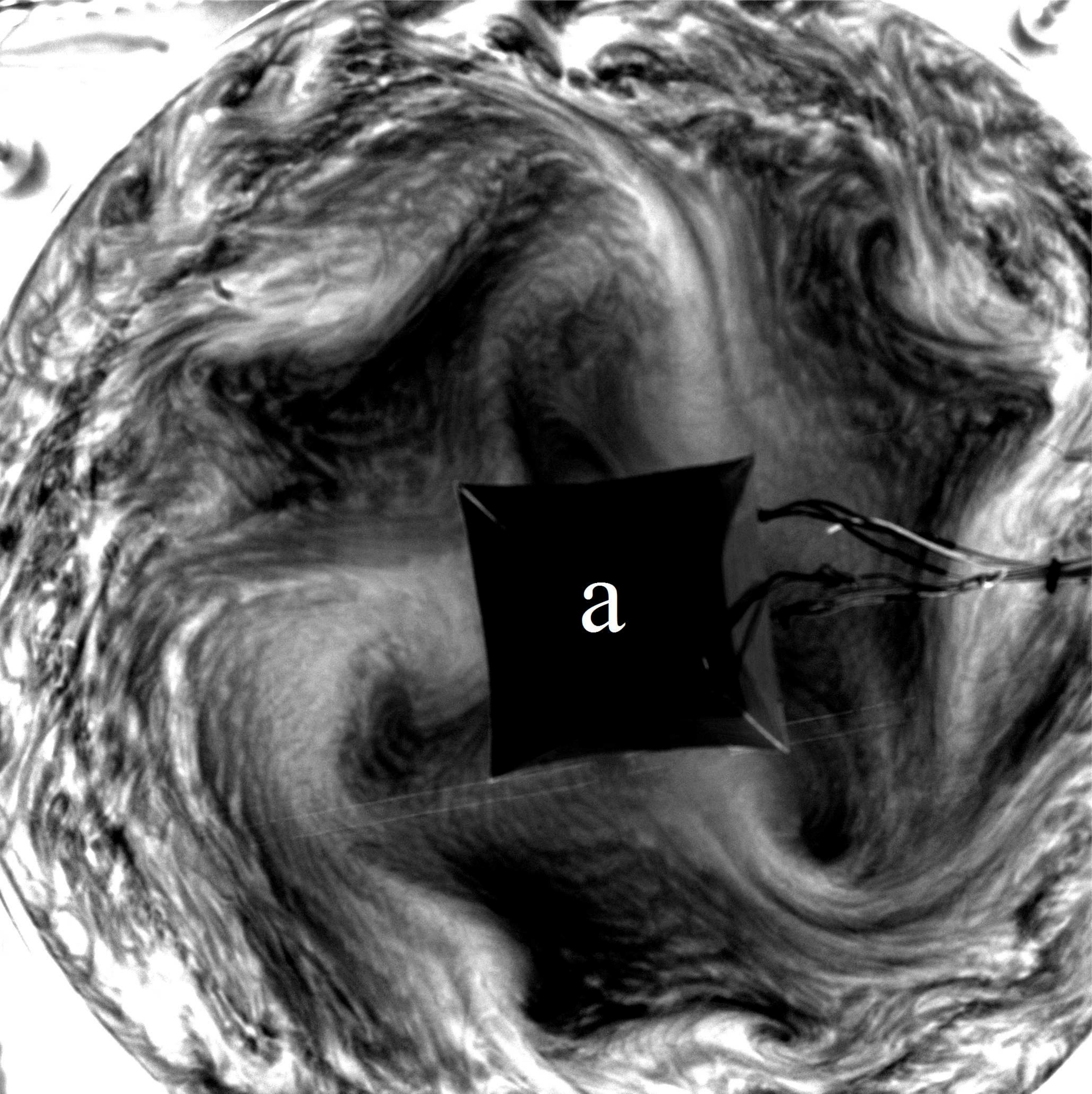} 
\includegraphics[width=0.33\linewidth]{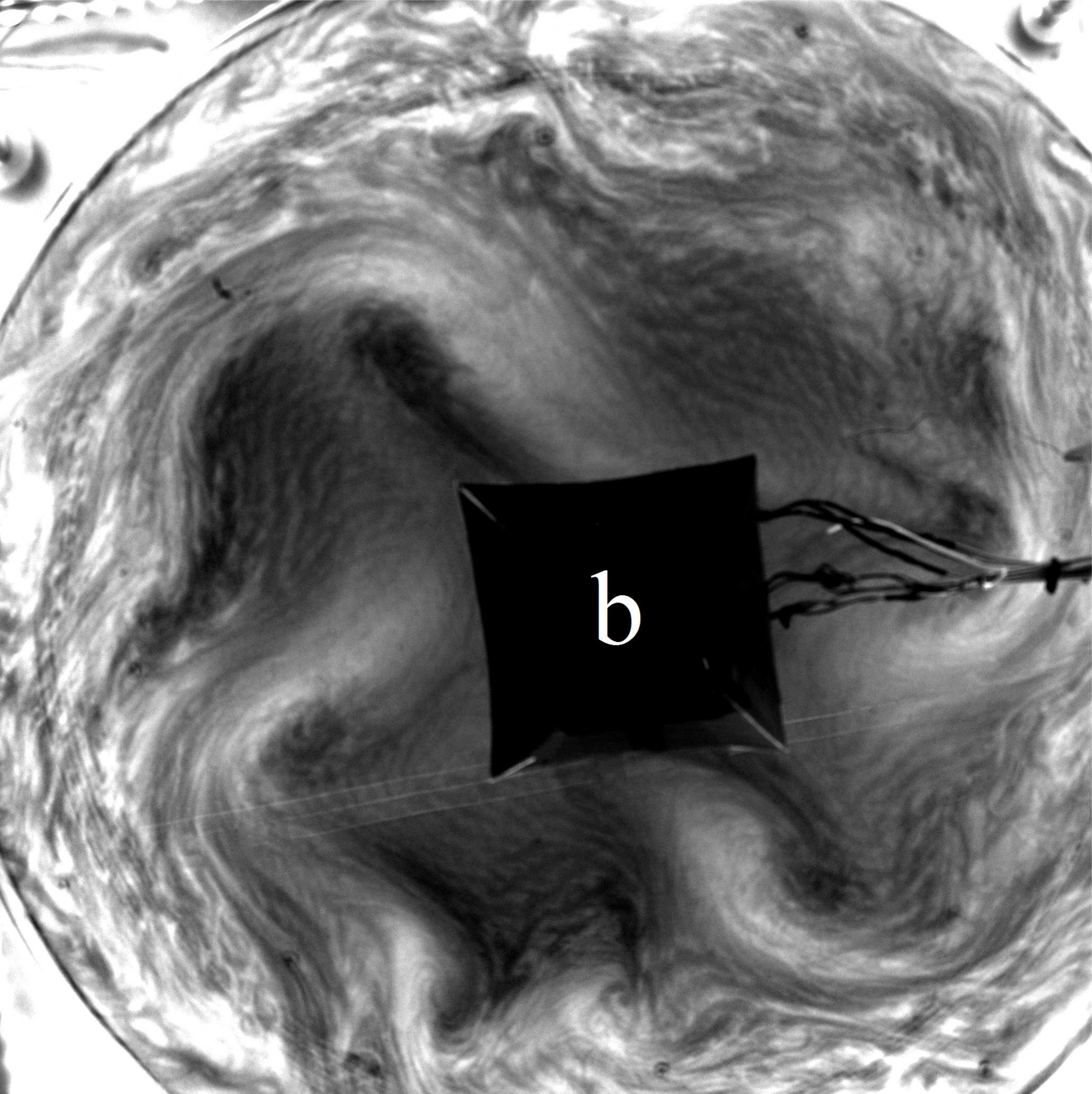}
\includegraphics[width=0.33\linewidth]{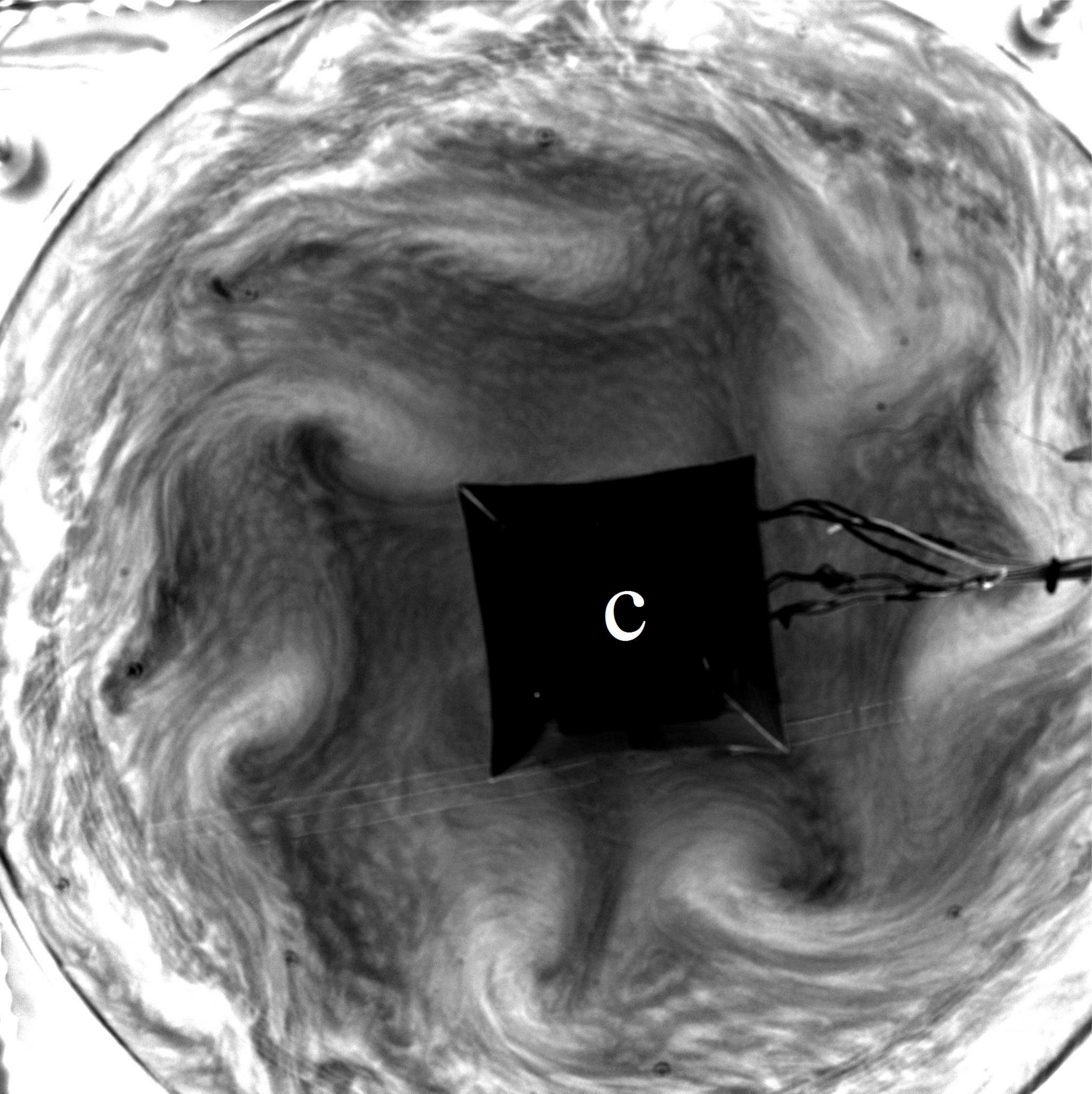} 
\caption{Variation of wave number during Exp.7, a -- m=3, b -- m=4, c -- m=7. Instanteneous images are shown.}
\label{Fig4}
\end{figure*}

\begin{figure*}
\includegraphics[width=0.33\linewidth]{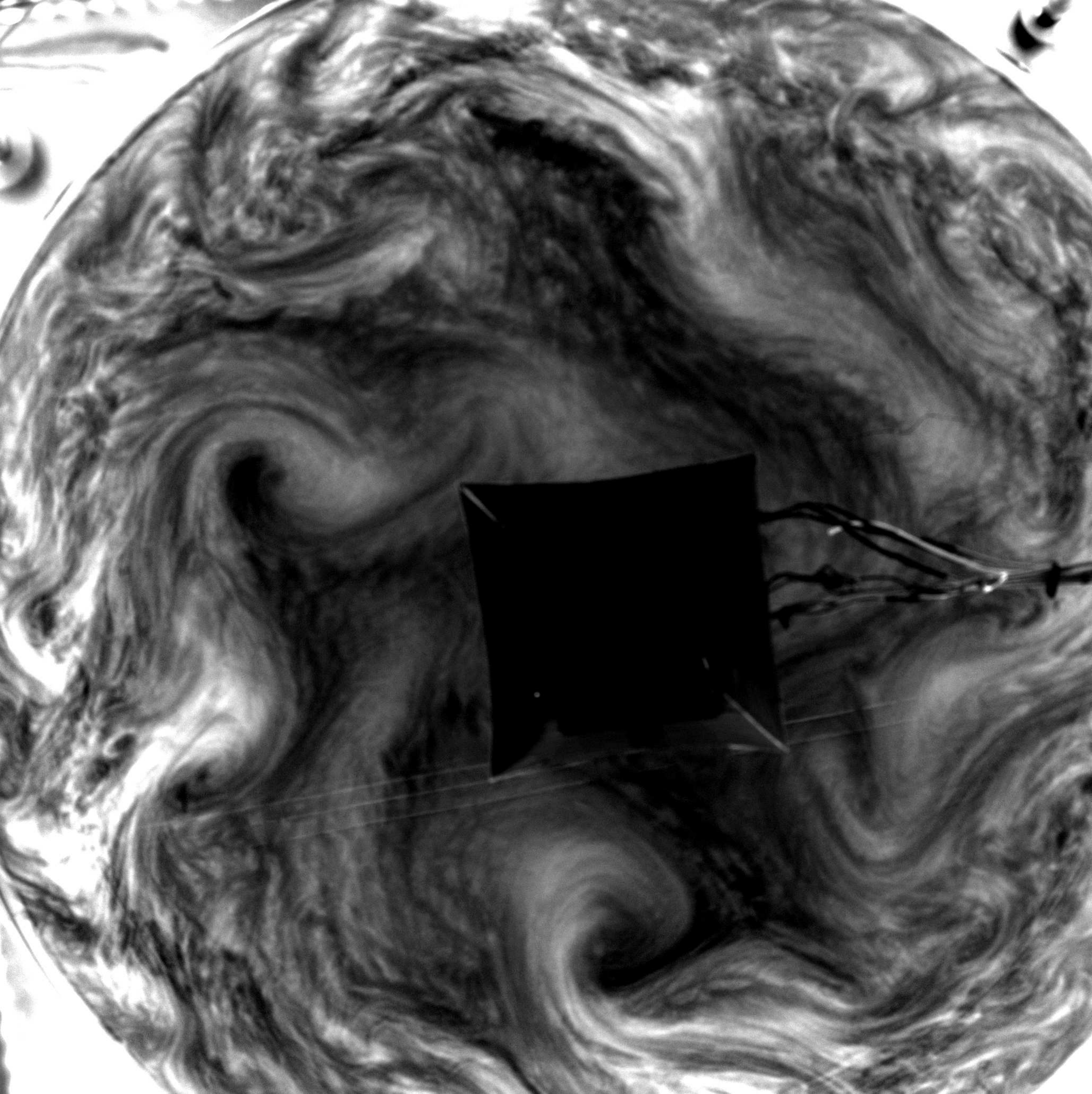} 
\includegraphics[width=0.33\linewidth]{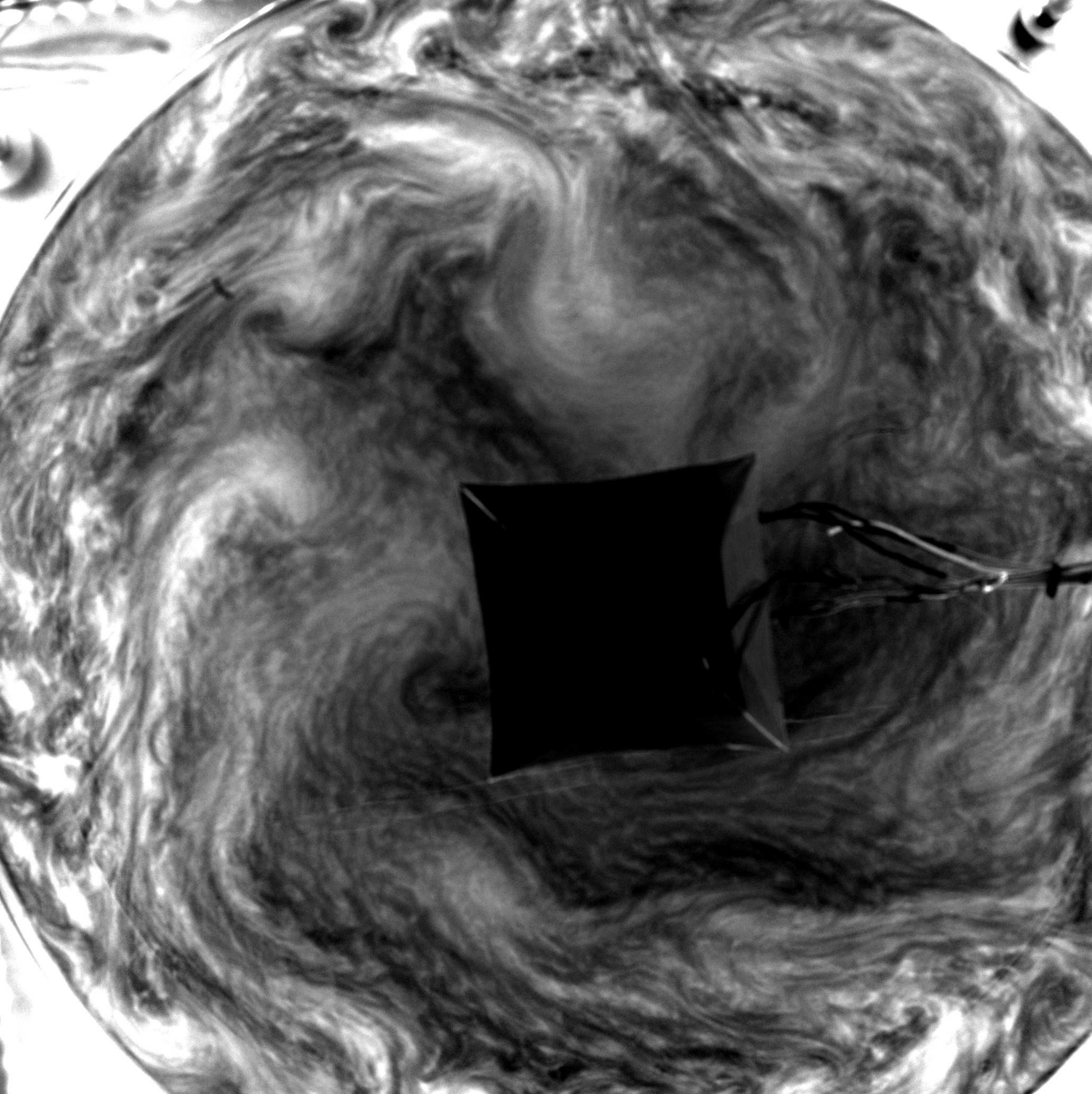}
\includegraphics[width=0.33\linewidth]{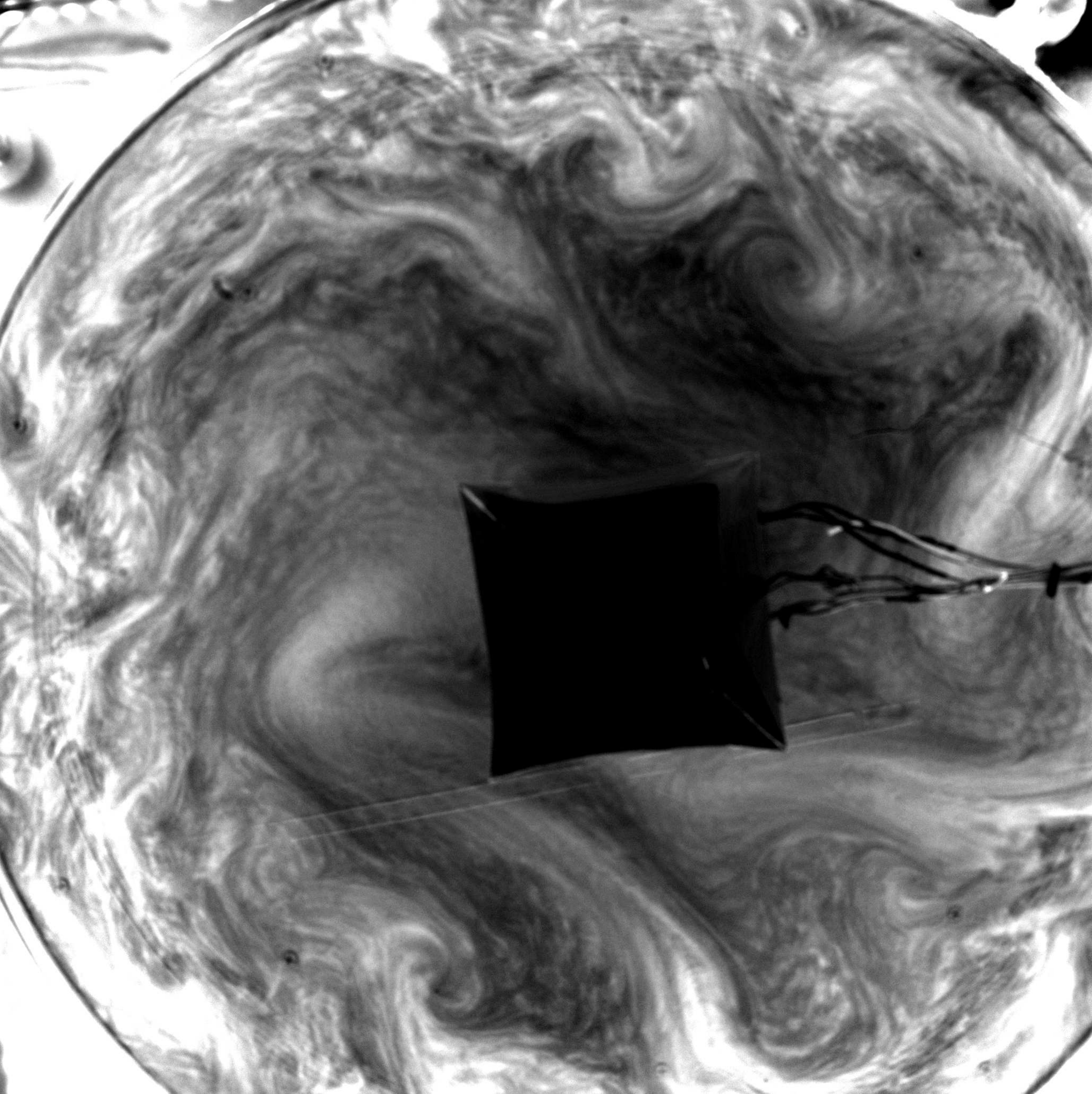} 

\caption{Typical flow patterns in irregular wave regime (Exp.8a).}
\label{Fig5}
\end{figure*}

\begin{figure*}
\center{\includegraphics[width=0.4\linewidth]{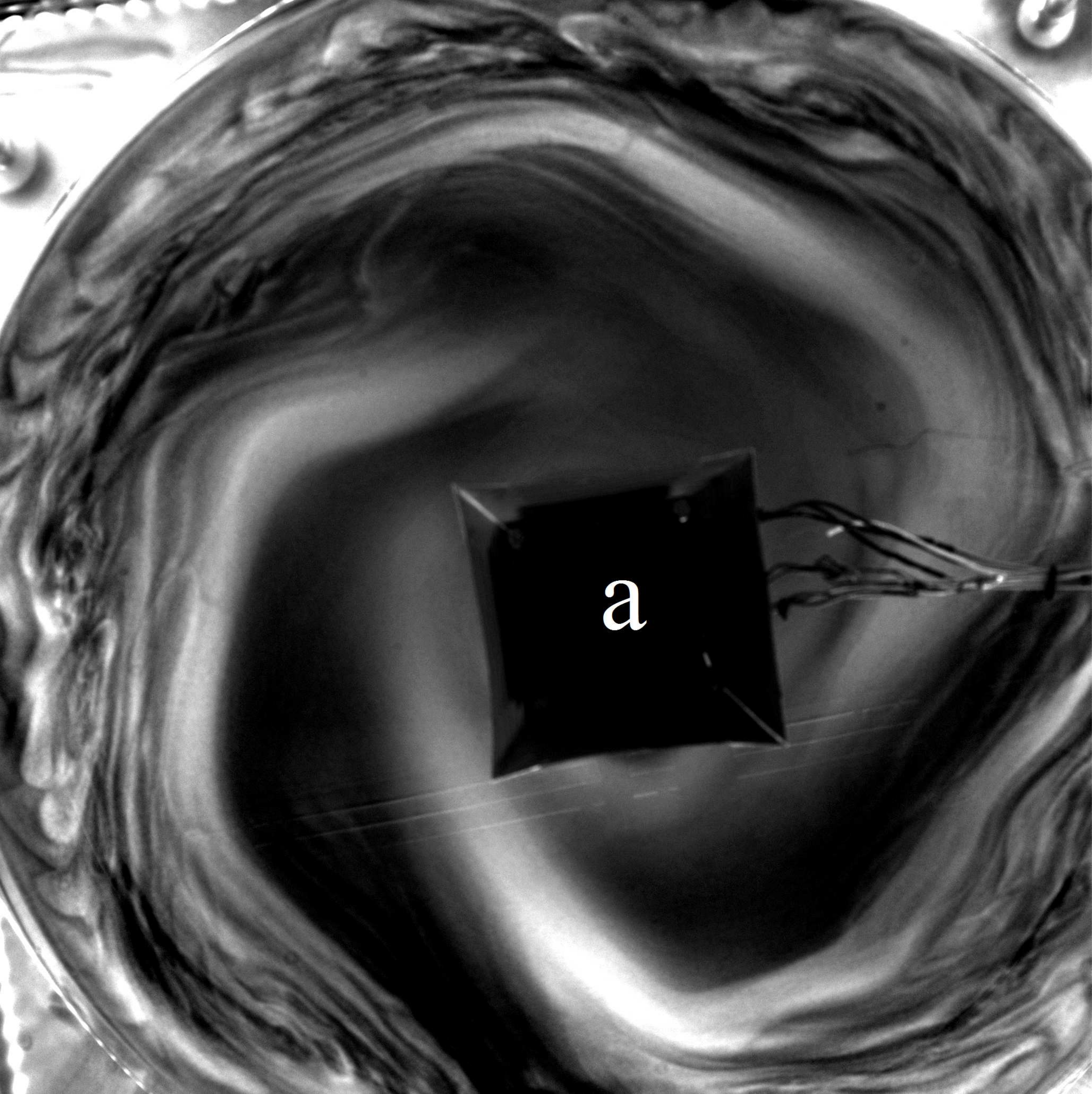} 
\includegraphics[width=0.4\linewidth]{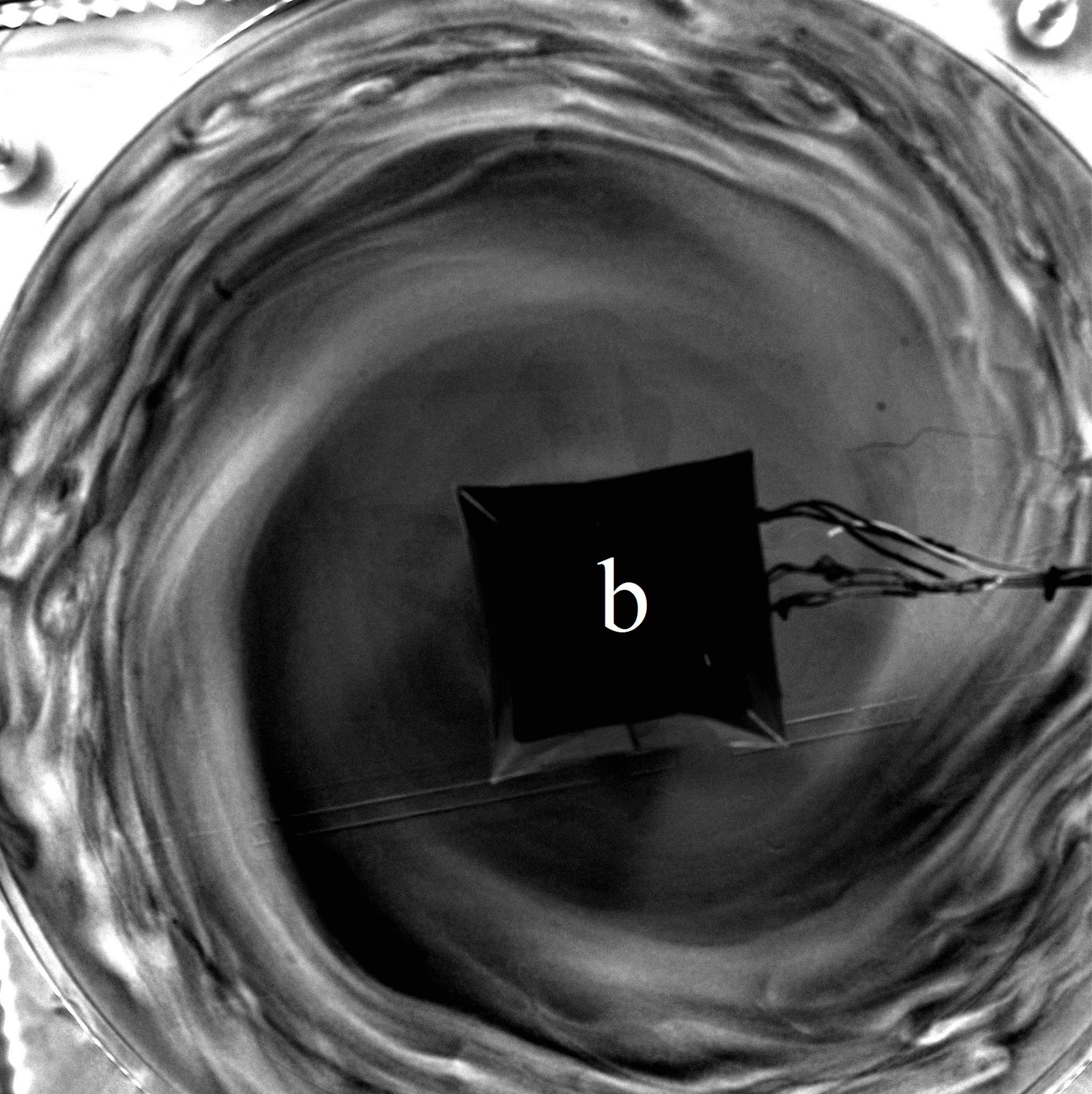}}
\caption{Regular baroclinic waves at low $Ro_T$, a -- Exp.8b, b -- Exp8c. Instanteneous images are shown.}
\label{Fig6}
\end{figure*}

The increase in rotation rate results in increase of Taylor number and decrease in thermal Rossby number. In our experiments in agreement with \citep{fultz1959studies,read2014general,scolan2017rotating} it leads to the instability of the axisymmetric flow. The instanteneous images in Fig.~\ref{Fig3} illustrate transition from Hadley-like regime to the regime with baroclinic waves.  There is an important difference between the flow structure and dynamics in the presented experiments and in a classical annulus configuration \citep{read2014general} and a new one proposed in \citep{scolan2017rotating}. In a classic configuration, there is an area of parameters on a map $Ra_T-Ta$ where steady regular waves are formed. Horizontal heating and cooling \citep{scolan2017rotating} decrease the area where steady waves exist. Here, in regimes with evident baroclinic waves, as in Fig.~\ref{Fig3}c, the flow is constantly evolving, which leads to the changes in the number, amplitude, and shape of the waves. The movies, which illustrate this temporal flow evolution are presented in Supplementary materials.  A further increase in rotation rate leads to the more complex flow structure and dynamics. Fig.~\ref{Fig4} shows variation in the number of waves from 3 to 7 during one experimental run (Exp.7). The time of living  of the regular structures strongly variates from several to tens of rotation periods. This regime with regular but unsteady waves is sensitive to the control parameters, and relatively small variation in $Ro_T$ and $Ta$ (Exp. 8a, Fig.~\ref{Fig5}) may lead to the irregular flow structure. Decreasing of $Ro_T$ by decreasing in the heating power ($P_h$=24.5 Wt, for Exp.8b)  results in the regularization of the flow (Fig.~\ref{Fig6}).  The baroclinic waves with the main mode $m=3$ are formed. Further decrease of $Ro_T$ ($P_h$=8 Wt,  Exp.8c) leads to the weakening of baroclinic waves and we can expect the achievement of the lower symmetrical regime \citep{mason1975baroclinic} with a subsequent decrease in $Ro_T$.

\subsection{Mode analysis}

In the previous section we provide a qualitative description of the flow structure and its evolution with the variation of the control parameters. Fourier decomposition of the brightness $I$ field of instantaneous grey scale images can provide valuable information about the modes of baroclinic waves and their temporal behaviour. The circled area in the mid-radius is used for Fourier decomposition. The Fourier components $A_n$ , $B_n$ and energy of Fourier modes $E_n$ are given by :

\begin{align}
A_n(t)=\frac{1}{L}\sum_{k} I_k(t)\cos(2n\pi k/L), \\
B_n(t)=\frac{1}{L}\sum_{k} I_k(t)\sin(2n\pi k/L), \\
E_n(t)={A_n(t)}^2+{B_n(t)}^2, 
\end{align}
where $t$ is a time, $k$ is an angle, $I_k(t)$ is a brightness value averaged over 80x80 pixels area at mid-radius circle,  $L=360$.

\begin{figure}
\center{\includegraphics[width=0.8\linewidth]{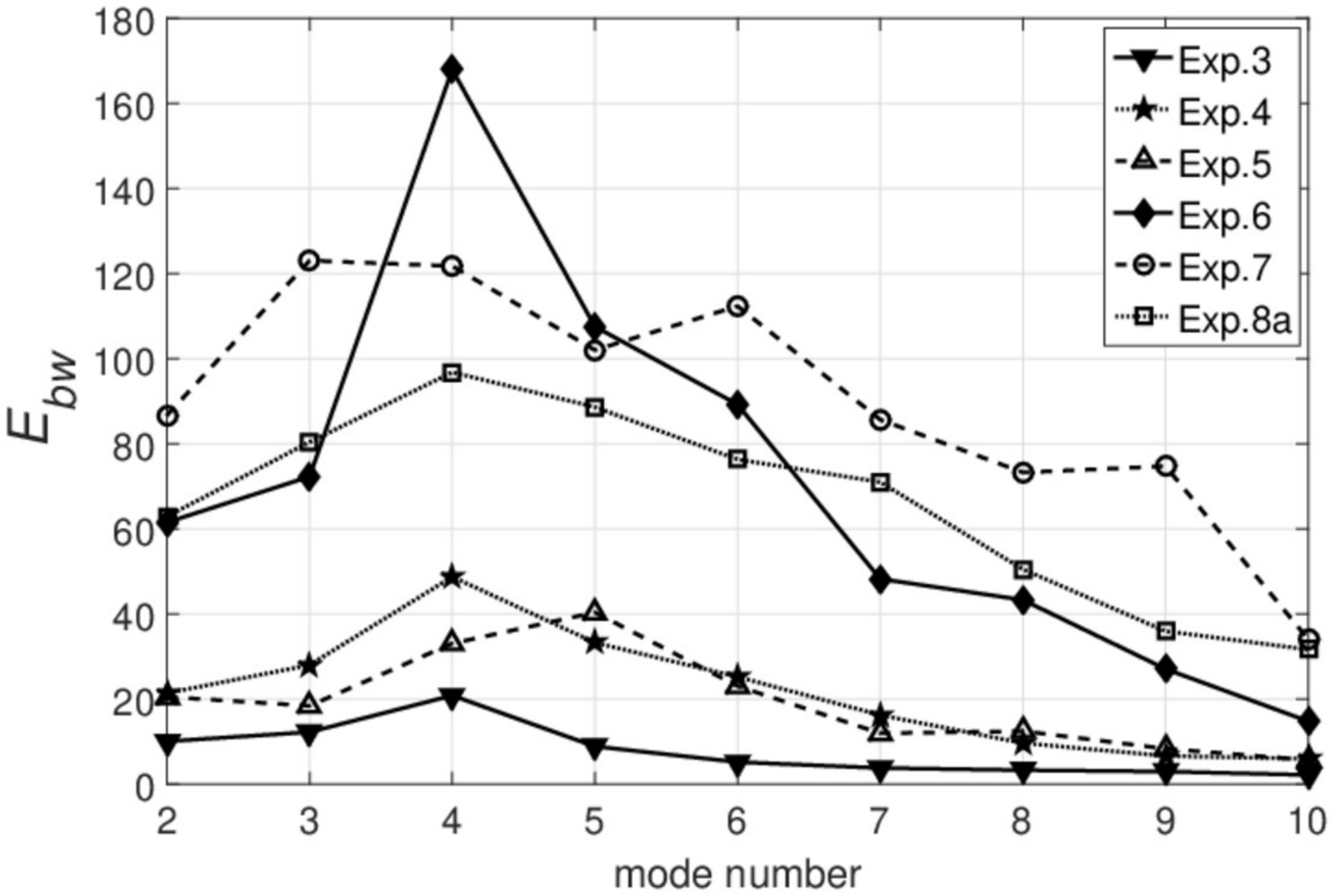}} 
\caption{Energy (in arbitrary units) of different modes of baroclinic waves. }
\label{Fig7}
\end{figure}

\begin{figure*}
\center{\includegraphics[width=1\linewidth]{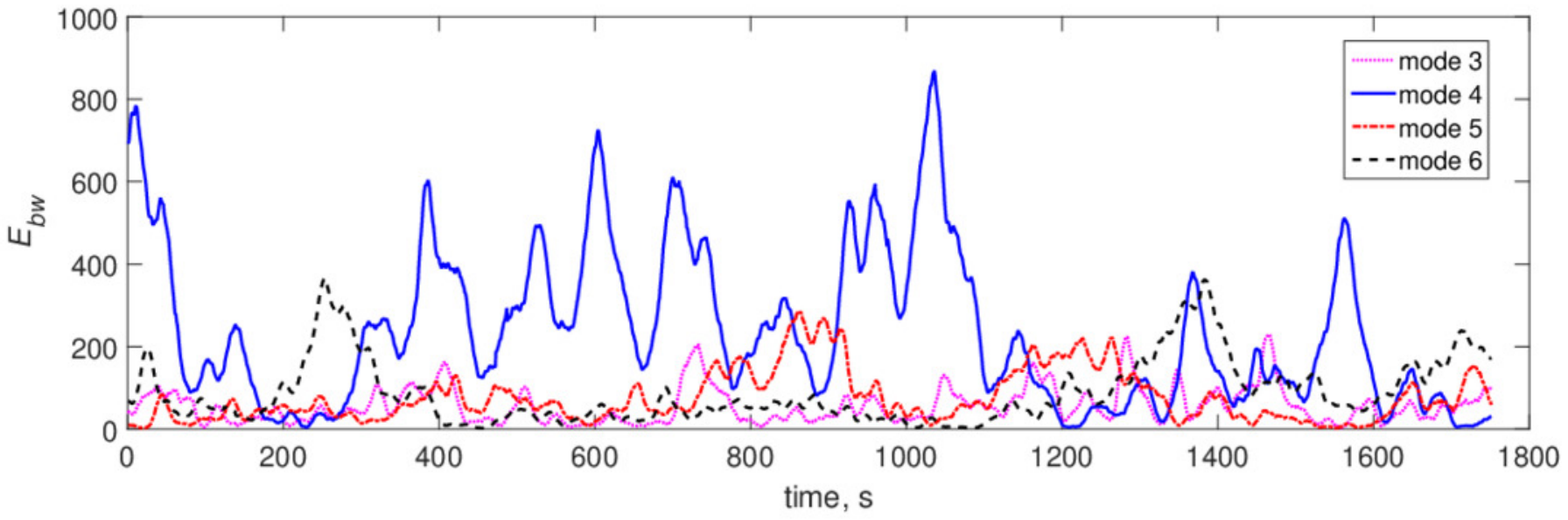}} a)\\
\center{\includegraphics[width=1\linewidth]{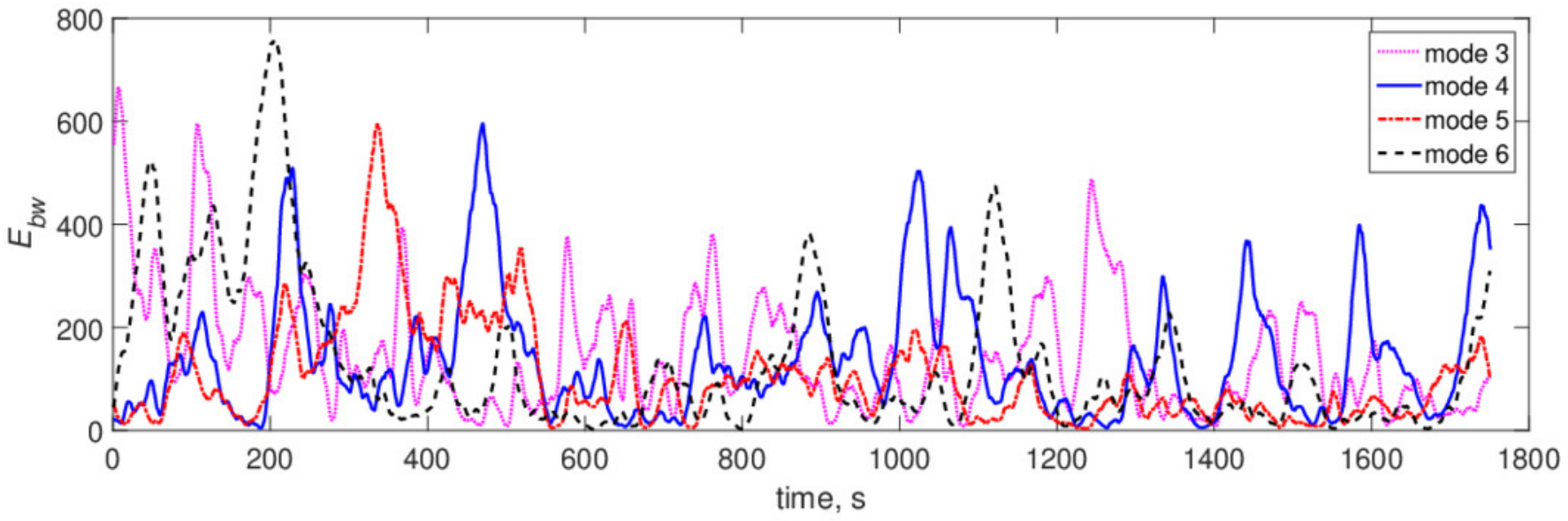}} b)\\
\center{\includegraphics[width=1\linewidth]{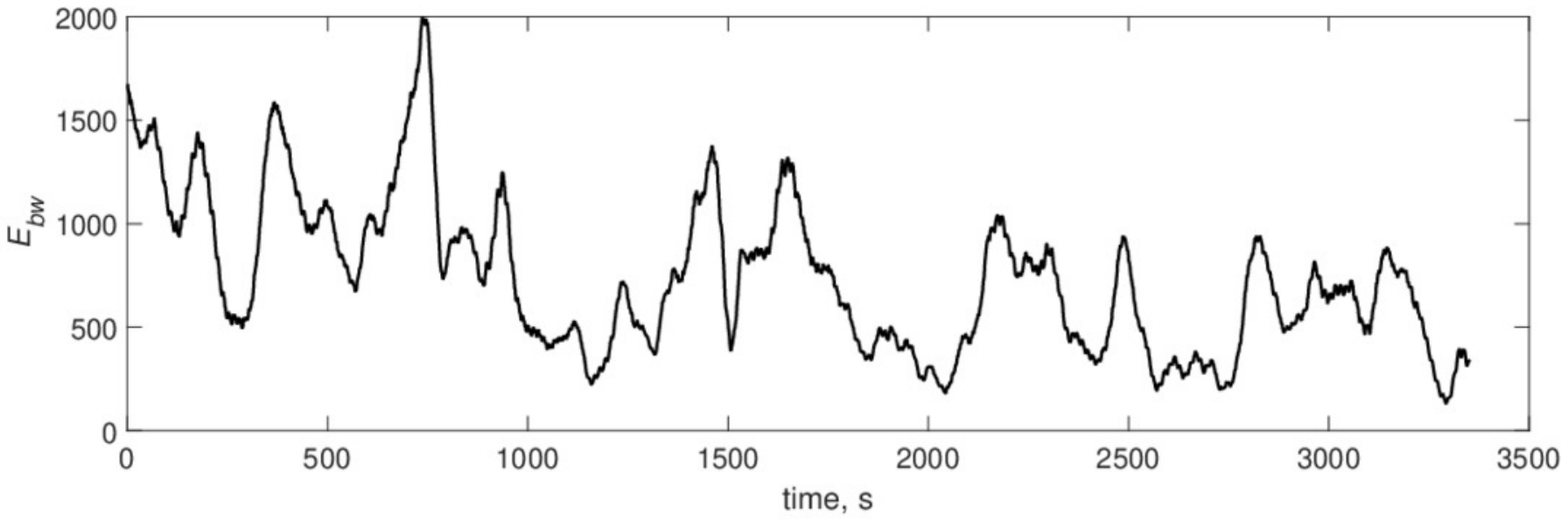}} c)\\
\caption{Time evolution of main wave modes, a -- Exp.6, b -- Exp.7, c -- Exp.8b ($m=3$). }
\label{Fig8}
\end{figure*}

\begin{figure*}
\center{\includegraphics[width=1\linewidth]{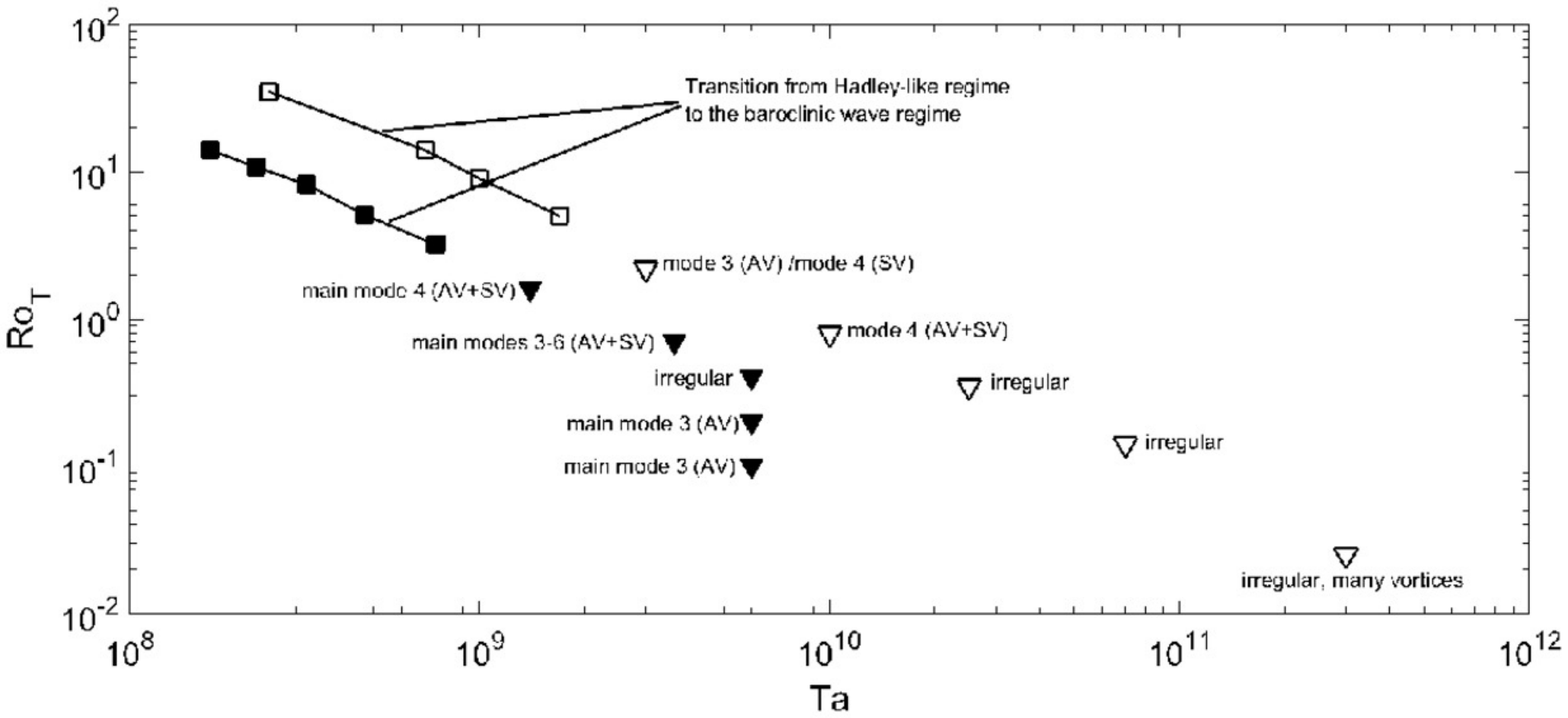}} 
\caption{Map of regimes. Results from \citep{scolan2017rotating} (open symbols) are plotted for comparison. AV -- amplitude vacillation, SV -- shape vacillation.}
\label{Fig9}
\end{figure*}

The energy of the main modes of baroclinic waves for a series of experiments with increasing rotation rates are presented in Fig.~\ref{Fig7}. These results are in a good agreement with our qualitative observations. Increasing the rotation rate leads to the instability of Hadley-like regime and the formation of baroclinic waves. For this configuration, the flow in a baroclinic regime is a combination of different modes of baroclinic waves. Only for the Exp.6 one dominant mode $m=4$ can be separated. The time evolution of different wave modes (Fig.~\ref{Fig8}) shows that even in Exp.6 the dominant mode is characterized by strong non-periodic fluctuations (Fig.~\ref{Fig8}a). The wave structures in Exp.7 is a superposition of several competing wave modes (Fig.~\ref{Fig8}b), which explains complex and irregular flow transformations. A substantial decrease in $Ro_T$ for the fixed $Ta$ results in a regular $m=3$ wave, but its amplitude also strongly and non-periodically fluctuates (Fig.~\ref{Fig8}c). The map of regimes in a plane $Ro_T -Ta$ is shown in Fig.~\ref{Fig9}, where results from \citep{scolan2017rotating} are plotted for comparison.

\section{Summary and conclusions}

A new shallow layer laboratory model of global atmosphere circulation is realized. The configuration of the model is similar to the dishpan configuration in \citep{fultz1959studies} but with some important differences. The heater in the shape of a rim is shifted from the sidewall. It provides an anticyclonic belt in the upper layer near the sidewall, modeling easterly winds and decreasing the influence of the sidewall on the flow formation. We use low-viscous silicon oil instead of water to avoid complex effects provided by the formation of a thin film of a surface-active substance on the open surface. The transformation of the flow structure with the increase in the rotation rate generally agrees with previous results (see regime diagram in \cite{scolan2017rotating}). The flow transforms from the Hadley-like regime to the baroclinic wave regime through transitional states. The decrease in the thermal Rossby number for the fixed value of Taylor number results in the regularization of the baroclinic waves.

The main difference between presented and classical annulus configurations is the absence of the steady waves. All wave regimes, even with regular wave structures, are characterized by strong non-periodic fluctuations. The observed baroclinic wave structures are a combination of temporarily evolving different baroclinic modes. We can assume that more stochastic and irregular flows in the presented configuration can be explained by the specific realization of the heating. The rim heating results in the formation of multiple plumes and non-homogeneous temperature distribution in the azimuthal direction. In a rotating layer, the structure of convective flow in the periphery becomes more non-uniform due to the formation of vortices at the border between anticyclonic and cyclonic flows. This strongly non-homogeneous azimuthal distribution of warm fluid results in the excitation of baroclinic waves of different modes, unlike the case when the heating is provided by isothermal sidewall. A similar result (the absence of steady waves) was obtained in \citep{scolan2017rotating} with the heater and cooler at horizontal boundaries. This proves that the spatial distribution of heating and cooling, and their location, is of primary importance for the baroclinic wave stability. Another important aspect is an aspect ratio of the model. Its variation leads to a strong shift of the areas with different regimes in the diagram $Ro_T -Ta$. 

Finally, we can conclude that the presented model provides the formation of atmospheric-like flows, characterized by complex temporal behaviour and can serve as an efficient tool for studying different aspects of global atmospheric circulation.  

\section{Acknowledgements}
\label{acknow}
This research was supported by Russian Science Foundation grant RSF-22-21-00572 (https://rscf.ru/project/22-21-00572/ ).

\markboth{\rm A. SUKHANOVSKI AND E. POPOVA}{\rm GEOPHYSICAL \&  ASTROPHYSICAL FLUID DYNAMICS}

\bibliographystyle{gGAF}
\bibliography{GGAF-2022-0001-Sukhanovskii-Popova}
\markboth{\rm A. SUKHANOVSKI AND E. POPOVA}{\rm GEOPHYSICAL \&  ASTROPHYSICAL FLUID DYNAMICS}
\vspace{36pt}


\end{document}